\documentclass[twocolumn,preprintnumbers,amsmath,amssymb]{revtex4}

\usepackage{graphicx}
\usepackage{dcolumn}
\usepackage{bm}
\usepackage{longtable}

\usepackage{color}

\begin{document}


\title{Projecting human development and CO$_2$ emissions}

\author{Lu{\'i}s Costa}
\email{luis.costa@pik-potsdam.de}
\author{Diego Rybski}
\email{ca-dr@rybski.de}
\author{J\"urgen P. Kropp}
\email{kropp@pik-potsdam.de}
\affiliation{
Potsdam Institute for Climate Impact Research (PIK),
P.O. Box 60 12 03,
14412 Potsdam,
Germany
}

\date{\today}



\maketitle

\tableofcontents

\newpage




\section{Synopsis}
This is the Supporting Information (SI) to our 
manuscript \emph{A Human Development Framework for CO$_2$ Reductions}.

We estimate cumulative CO$_2$ emissions during the period $2000$ to $2050$ from
developed and developing countries based on the empirical relationship between
CO$_2$ per capita emissions (due to fossil fuel combustion and cement
production) and corresponding HDI. We choose not to include emissions from land
use and other greenhouse gases since they were found not to be strongly
correlated with personal consumption and national carbon intensities
\cite{Chakravarty2010}. In addition, data of past CO$_2$ emissions from land
use is uncertain due to the lack of historical data of both former ecosystem
conditions and the extent of subsequent land use \cite{Rhemtulla2008}. 

In order to project per capita emissions of individual countries we make three
assumptions which are detailed below. 
First, we use logistic regressions to fit and extrapolate the HDI on a country
level as a function of time. This is mainly motivated by the fact that the HDI
is bounded between~$0$ and~$1$ and that it decelerates as it approaches~$1$.
Second, we employ for individual countries the correlations between CO$_2$ per
capita emissions and HDI in order to extrapolate their emissions. This is an
ergodic assumption, i.e. that the process over time and over the statistical
ensemble is the same. Third, we let countries with incomplete data records
evolve similarly as their close neighbors (in the emissions-HDI plane, 
see Fig.~1 in the main text) with complete time series of CO$_2$ per capita
emissions and HDI. Country--based emissions estimates are obtained by
multiplying extrapolated CO$_2$ per capita values by population numbers of three
scenarios extracted from the Millennium Ecosystem Assessment report
\cite{Alcamo2005}. 

Finally, we propose a reduction scheme, where countries with an HDI 
above the development threshold reduce their per capita CO$_2$ emissions 
with a rate that is proportional to their HDI. 
We estimate the minimum proportionality constant so that the global emissions 
by $2050$ meet the $1000$\,Gt limit.

\section{Data}
The analyzed data consists of
Human Development Index (HDI),
CO$_2$ emissions per capita values, and
Population numbers.
In all cases the aggregation level is country scale.
Both the HDI and the CO$_2$ data is incomplete,
i.e. the values of some countries or years are missing.
In addition, the set of countries with HDI or CO$_2$ data does not match
$100$\% with the set of countries with population data
(see Sec.~\ref{subsec:limitations}).

\subsection{Human Development Index (HDI)}
\label{ssec:hdi}
The Human Development Index is provided by the
United Nation Human Development Report 2009 and
covers the period $1980$ to $2007$
(in steps of 5 years until $2005$, plus the years~$2006$ and~$2007$).
The data is available for download \cite{hdidata} and
is documented \cite{hdidocu}.

The HDI is intended to reflect three dimensions of human development:
(i) a long and healthy life,
(ii) knowledge, and
(iii) a decent standard of living.
In order to capture the dimensions, four indicators are used:
life expectancy at birth for "a long and healthy life",
adult literacy rate and gross enrollment ratio (GER) for "knowledge", and
GDP per capita (PPP US\$) for "a decent standard of living".
Each index is weighted with $1/3$ whereas the "adult literacy index"
contributes $2/3$ to the education index (knowledge) and
gross enrollment index $1/3$:
\begin{eqnarray}
d_{i,t}&=&
\frac{1}{3}\left(\frac{LE_{i,t}-25}{85-25}\right)+ \nonumber \\
& & \quad\frac{1}{3}\left(\frac{2}{3}\frac{AL_{i,t}}{100}+\frac{1}{3}\frac{GE_{i,t}}{100}\right)+ \nonumber \\
& & \quad\frac{1}{3}\left(\frac{\log{GDP_{i,t}}-\log{100}}{\log{40000}-\log{100}}\right) \\
&=& \frac{1}{3}d_{i,t}^\text{life expectancy}+
\frac{1}{3}d_{i,t}^\text{education}+
\frac{1}{3}d_{i,t}^\text{GDP}
\enspace ,
\end{eqnarray}
where $LE_{i,t}$ is the life expectancy, 
$AL_{i,t}$ the adult literacy, 
$GE_{i,t}$ the gross enrollment, and 
$GDP_{i,t}$ the GDP per capita (PPP US\$) \cite{UNDP2008}, 
$d_{i,t}^\text{life expectancy}$, 
$d_{i,t}^\text{education}$, and 
$d_{i,t}^\text{GDP}$ 
denote the corresponding indices.
An illustrative diagram can be found in \cite{UNDP2008}.
The components are studied individually in 
Sec.~\ref{ssubsec:components}.


\subsection{CO$_2$ emissions per capita}
The data on CO$_2$ emissions per capita is provided by the
World Resources Institute (WRI) $2009$ and
covers the years $1960$-$2006$.
The CO$_2$ emissions per capita are given in units of tons per year.
It is available for download \cite{co2data} and
is documented \cite{co2docu}.

Carbon dioxide (CO$_2$) is transformed and released during
combustion of solid, liquid, and gaseous fuels \cite{co2docu2}.
In addition, CO$_2$ is emitted as cement is calcined to
produce calcium oxide.
The data does include emissions from cement production but
estimates of gas flaring are included only from $1980$ to the present.
The CO$_2$ emission values do not include emissions from land
use change or emissions from bunker fuels used in international
transportation \cite{co2docu2}.


\subsection{Population}
\label{ssec:datapop}
Population projections are provided by
the Millennium Ecosystem Assessment Report 2001
and cover the period $2000$ to $2100$ in steps of $5$~years
(we only make use of the data until $2050$).
The data is available for download \cite{popdata} and is
documented \cite{popdocu}.
We use the scenarios
Adaptive Mosaic~(AM),
TechnoGarden~(TG), and
Global Orchestration~(GO).
We found minimal differences in our results using the
Order from Strength~(OS) scenario 
and therefore disregard it. 
In short:

\begin{itemize}
\item
The Adapting Mosaic scenario is based on a fragmented world resulting from
discredited global institutions. It involves the rise of local ecosystem
management strategies and the strengthening of local institutions
\cite{popdocu}.
\item
The TechnoGarden scenario is based on a globally connected world relying
strongly on technology as well as on highly managed and often-engineered
ecosystems to provide needed goods and services.
\item
The Global Orchestration scenario is based on a worldwide connected society
in which global markets are well developed. Supra-national institutions
are well established to deal with global environmental problems.
\end{itemize}


\subsection{Notation}
\label{subsec:notation}
For a country~$i$ at year~$t$ we use the following quantities:
\begin{itemize}
\item
Human Development Index (HDI): $d_{i,t}$
\item
CO$_2$ emissions per capita: $e^{(\rm c)}_{i,t}$\\
in tons/(capita{\enspace}year)
\item
CO$_2$ emissions: $e_{i,t}$\\
in tons/year
\item
cumulative CO$_2$ emissions: $E_{i,t}$\\
in tons
\item
population: $p_{i,t}$
\end{itemize}

\section{Extrapolating CO$_2$ emissions}
\label{sec:co2project}
In this section we detail which empirical findings and assumptions are
used to extrapolate per capita emissions of CO$_2$ and HDI values
in a Development As Usual (DAU) approach.
The projections are performed statistically, i.e. extrapolating regressions.
Our approach is based on $3$~assumptions:
\begin{enumerate}
\item
The Human Development Index, $d_{i,t}$, of all countries evolves
in time following logistic regressions
(Sec.~\ref{subsec:extrahdi}).
\item
The Human Development Index and the logarithm of the CO$_2$ emissions
per capita, $e^{(\rm c)}_{i,t}$, are linearly correlated
(Sec.~\ref{subsec:extraco2}).
\item
The changes of $d_{i,t}$ and $e^{(\rm c)}_{i,t}$ are correlated among
the countries, i.e. countries with similar values comprise similar changes
(Sec.~\ref{subsec:extramis}).
\end{enumerate}

By Development As Usual we mean that the countries behave as in the past,
with respect to these $3$~points.
In particular, past behavior may be extrapolated to the future.

It is impossible to predict how countries will develop and
how much CO$_2$ will be emitted per capita.
Accordingly, we are not claiming that the calculated extrapolations are
predictions.
We rather present a plausible approach which is supported by the
development and the emissions per capita in the past.
We provide the estimates consisting of projected HDI and emission values
as supplementary material.

\subsection{Extrapolating Human Development Index (HDI)}
\label{subsec:extrahdi}

We elaborate the evolution of HDI values following a
logistic regression \cite{HosmerL2000}.
This choice is supported by the fact that the HDI is bounded to
$0\le d_{i,t}\le 1$ and that the high HDI countries develop slowly.
Therefore, we fit for each country separately
\begin{equation}
\label{eq:hdilogreg}
\tilde{d}_{i,t} = \frac{1}{1+{\rm e}^{-a_it+b_i}}
\end{equation}
to the available data (obtaining the parameters~$a_i$ and~$b_i$),
whereas we only take into account those countries for which we
have at least $4$~measurement points, which leads to regressions for
$147$~countries out of $173$ in our data set.
Basically, $a_i$ quantifies how fast a country develops and
$b_i$ represents when the development takes place. 
Figure~2 in the main paper depicts a collapse 
(see e.g. \cite{MalmgrenSCA2009}) 
of the past HDI as obtained from the logistic regression. 
It illustrates how the countries have been developing in the scope of
this approach.

\begin{table*}
\begin{tabular}{| l || c | c | c | c | c | c | c | c | c | c |}
\hline
	&	2007	&	2012	&	2017	&	2022	&	2027	&	2032	&	2037	&	2042	&	2047	 \\
	&	2011	&	2016	&	2021	&	2026	&	2031	&	2036	&	2041	&	2046	&	2051	 \\
\hline
\hline
Armenia	&	$\bullet$	&		&		&		&		&		&		&		&	\\
\hline
Colombia	&	$\bullet$	&		&		&		&		&		&		&		&	\\
\hline
Iran	&	$\bullet$	&		&		&		&		&		&		&		&	\\
\hline
Kazakhstan	&	$\bullet$	&		&		&		&		&		&		&		&	\\
\hline
Mauritius	&	$\bullet$	&		&		&		&		&		&		&		&	\\
\hline
Peru	&	$\bullet$	&		&		&		&		&		&		&		&	\\
\hline
Turkey	&	$\bullet$	&		&		&		&		&		&		&		&	\\
\hline
Ukraine	&	$\bullet$	&		&		&		&		&		&		&		&	\\
\hline
Azerbaijan	&		&	$\bullet$	&		&		&		&		&		&		&	\\
\hline
Belize	&		&	$\bullet$	&		&		&		&		&		&		&	\\
\hline
China	&		&	$\bullet$	&		&		&		&		&		&		&	\\
\hline
Dominican.Republic	&		&	$\bullet$	&		&		&		&		&		&		&	\\
\hline
El.Salvador	&		&	$\bullet$	&		&		&		&		&		&		&	\\
\hline
Georgia	&		&	$\bullet$	&		&		&		&		&		&		&	\\
\hline
Jamaica	&		&	$\bullet$	&		&		&		&		&		&		&	\\
\hline
Maldives	&		&	$\bullet$	&		&		&		&		&		&		&	\\
\hline
Samoa	&		&	$\bullet$	&		&		&		&		&		&		&	\\
\hline
Suriname	&		&	$\bullet$	&		&		&		&		&		&		&	\\
\hline
Thailand	&		&	$\bullet$	&		&		&		&		&		&		&	\\
\hline
Tonga	&		&	$\bullet$	&		&		&		&		&		&		&	\\
\hline
Tunisia	&		&	$\bullet$	&		&		&		&		&		&		&	\\
\hline
Algeria	&		&		&	$\bullet$	&		&		&		&		&		&	\\
\hline
Bolivia	&		&		&	$\bullet$	&		&		&		&		&		&	\\
\hline
Fiji	&		&		&	$\bullet$	&		&		&		&		&		&	\\
\hline
Honduras	&		&		&	$\bullet$	&		&		&		&		&		&	\\
\hline
Indonesia	&		&		&	$\bullet$	&		&		&		&		&		&	\\
\hline
Jordan	&		&		&	$\bullet$	&		&		&		&		&		&	\\
\hline
Sri.Lanka	&		&		&	$\bullet$	&		&		&		&		&		&	\\
\hline
Syrian.Arab.Republic	&		&		&	$\bullet$	&		&		&		&		&		&	\\
\hline
Turkmenistan	&		&		&	$\bullet$	&		&		&		&		&		&	\\
\hline
Viet.Nam	&		&		&	$\bullet$	&		&		&		&		&		&	\\
\hline
Cape.Verde	&		&		&		&	$\bullet$	&		&		&		&		&	\\
\hline
Egypt	&		&		&		&	$\bullet$	&		&		&		&		&	\\
\hline
Equatorial.Guinea	&		&		&		&	$\bullet$	&		&		&		&		&	\\
\hline
Guatemala	&		&		&		&	$\bullet$	&		&		&		&		&	\\
\hline
Guyana	&		&		&		&	$\bullet$	&		&		&		&		&	\\
\hline
Mongolia	&		&		&		&	$\bullet$	&		&		&		&		&	\\
\hline
Paraguay	&		&		&		&	$\bullet$	&		&		&		&		&	\\
\hline
Philippines	&		&		&		&	$\bullet$	&		&		&		&		&	\\
\hline
Kyrgyzstan	&		&		&		&		&	$\bullet$	&		&		&		&	\\
\hline
Nicaragua	&		&		&		&		&	$\bullet$	&		&		&		&	\\
\hline
Uzbekistan	&		&		&		&		&	$\bullet$	&		&		&		&	\\
\hline
Lao.People's.Democratic.Republic	&		&		&		&		&		&	$\bullet$	&		&		&	 \\
\hline
Morocco	&		&		&		&		&		&	$\bullet$	&		&		&	\\
\hline
Vanuatu	&		&		&		&		&		&	$\bullet$	&		&		&	\\
\hline
Botswana	&		&		&		&		&		&		&	$\bullet$	&		&	\\
\hline
India	&		&		&		&		&		&		&	$\bullet$	&		&	\\
\hline
Nepal	&		&		&		&		&		&		&	$\bullet$	&		&	\\
\hline
Bangladesh	&		&		&		&		&		&		&		&	$\bullet$	&	\\
\hline
Sao.Tome.and.Principe	&		&		&		&		&		&		&		&	$\bullet$	&	\\
\hline
Yemen	&		&		&		&		&		&		&		&	$\bullet$	&	\\
\hline
Bhutan	&		&		&		&		&		&		&		&		&	$\bullet$	\\
\hline
Ethiopia	&		&		&		&		&		&		&		&		&	$\bullet$	\\
\hline
Pakistan	&		&		&		&		&		&		&		&		&	$\bullet$	\\
\hline
Solomon.Islands	&		&		&		&		&		&		&		&		&	$\bullet$	\\
\hline
Uganda	&		&		&		&		&		&		&		&		&	$\bullet$	\\
\hline
\end{tabular}

\caption{
Periods during which countries are expected to pass the
HDI value of 0.8 according to the extrapolations.
The rows denote the countries and the columns denote periods of five years.
The transitions are indicated with $\bullet$.}
\label{tab:transitions}
\end{table*}

Based on the obtained parameters,~$a_i$ and~$b_i$,
we estimate the future HDI of each country
assuming similar development trajectories as in the past.
Table~\ref{tab:transitions} lists those countries which pass
$d_{i,t}=0.8$ \cite{hdidocu} until $2051$ and provides periods when this
is expected to happen according to our projections.
Further, we expect from the extrapolations that before $2021$
more people will be living in countries with HDI above $0.8$ (see main text)
than below.
In addition, until $2051$ around $85$\% will be living in countries with HDI
above $0.8$.

The logistic regression, Eq.~(\ref{eq:hdilogreg}), is in physics 
also known as Fermi-Dirac distribution.
It comprises three distinct points. 
The inflection point is located at $t=0$ and $d=0.5$ for $a_i=1$ and $b_i=0$. 
Two other distinct points are those of maximum or minimum curvature. 
They are located at $t=-\ln(2\pm\sqrt{3})$ and $d=(3\pm\sqrt{3})^{-1}$, 
i.e. $d\approx0.21 \vee d\approx 0.79$. 
Accordingly, from a geometrical point of view, $d^*=0.8$ is a 
reasonable threshold.
The approach of fitting logistic regressions to country data 
is also been used in other fields, see e.g. \cite{HuTWZ2011}.





\subsection{Estimating CO$_2$ emissions per capita}
\label{subsec:extraco2}


%
\begin{figure*}
\includegraphics[width=0.7\textwidth]{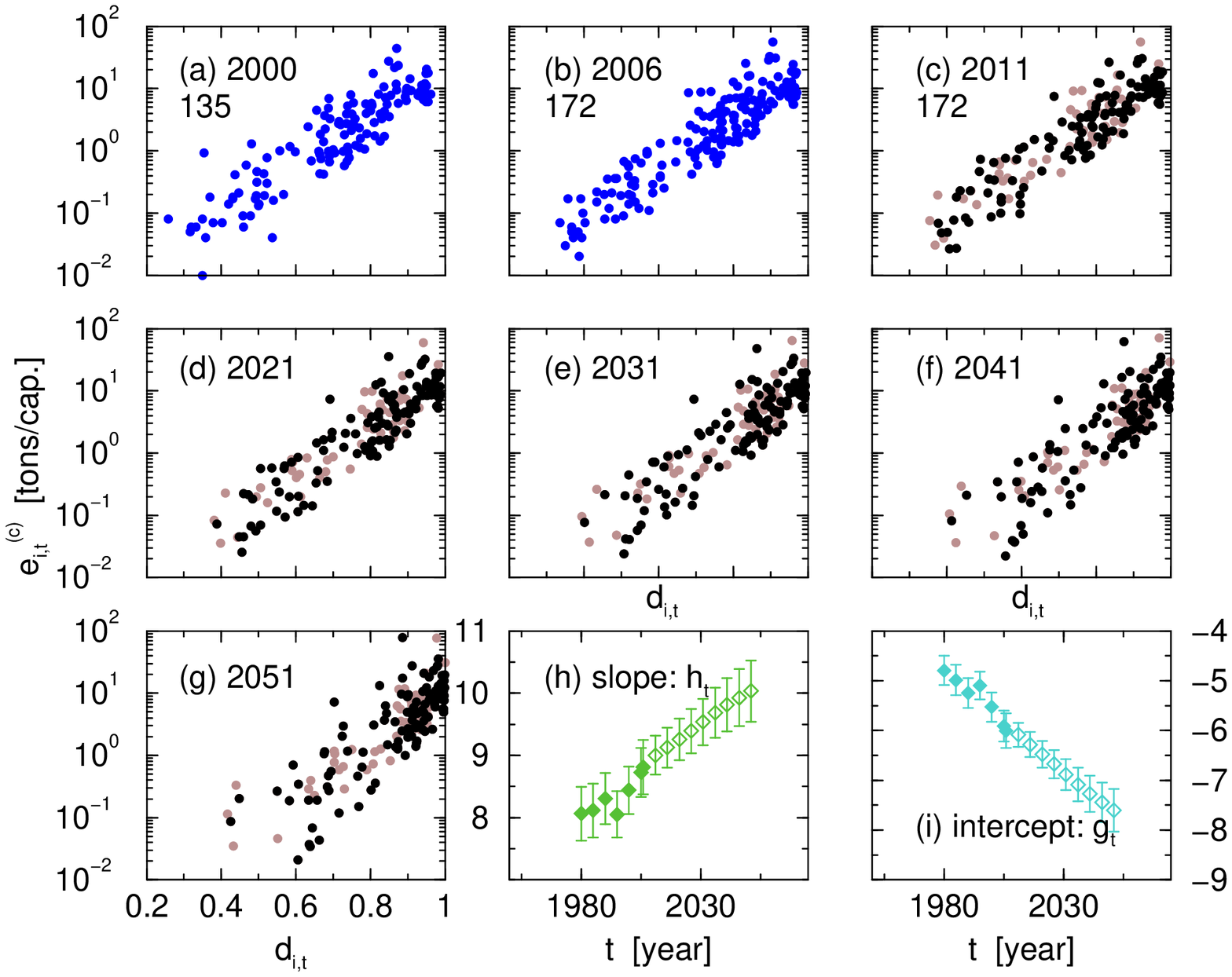}
\caption{
\label{fig:co2hdiexpmis}
Correlations between CO$_2$ emissions per capita and HDI.
Panels~(a) to~(g) are cross-plots in semi-logarithmic representation,
where each filled circle represents a country,
for past years
(a) $2000$: $135$ countries and
(b) $2006$: $172$ countries, as well as extrapolated
(c) to~(g) $2011$-$2051$ ($172$ countries each).
The brownish circles represent those countries,
which due to missing data have been estimated 
assuming correlations in the changes of $d_{i,t}$ as well as 
$e^\text{(c)}_{i,t}$ (see Sec.~\ref{subsec:extramis}).
Panels (h) and (i) show how the parameters~$h_t$ and~$g_t$ evolve in
time (the open symbols are obtained from the extrapolated values of all
countries).
Both parameters are based on only those $71$~countries
providing HDI and CO$_2$ values for all years $1980$, $1985$, 
$1990$, $1995$,
$2000$, $2005$, $2006$. 
The qualitative agreement of~$h_t$ and~$g_t$ between past and extrapolated
supports the plausibility of the presented approach.
The error bars are given by the standard errors.
The panels~(h) and~(i) are the same as Fig.~3(c) and~(d) in the main text.
}
\end{figure*}
%

In Figure~1 of the main text we find among the ensemble of countries
correlations between the HDI, $d_{i,t}$, and the CO$_2$ emissions per capita,
$e^{(\rm c)}_{i,t}$ (see also Fig.~\ref{fig:co2hdiexpmis}).
We apply the exponential regression
\begin{equation}
\label{eq:co2hdiexp}
\hat{e}^{(\rm c)}_{i,t} = {\rm e}^{h_{t} d_{i,t} + g_{t}}
\end{equation}
to the country data by linear regression \cite{MasonGH1989} through
$\ln e^{(\rm c)}_{i,t}$ versus $d_{i,t}$ for fixed years~$t$ and
obtain the parameters~$h_t$ and~$g_t$ as displayed in the
panels~(c) and~(d) of Fig.~3 in the main text, respectively.


%
\begin{figure*}
\includegraphics[width=0.7\textwidth]{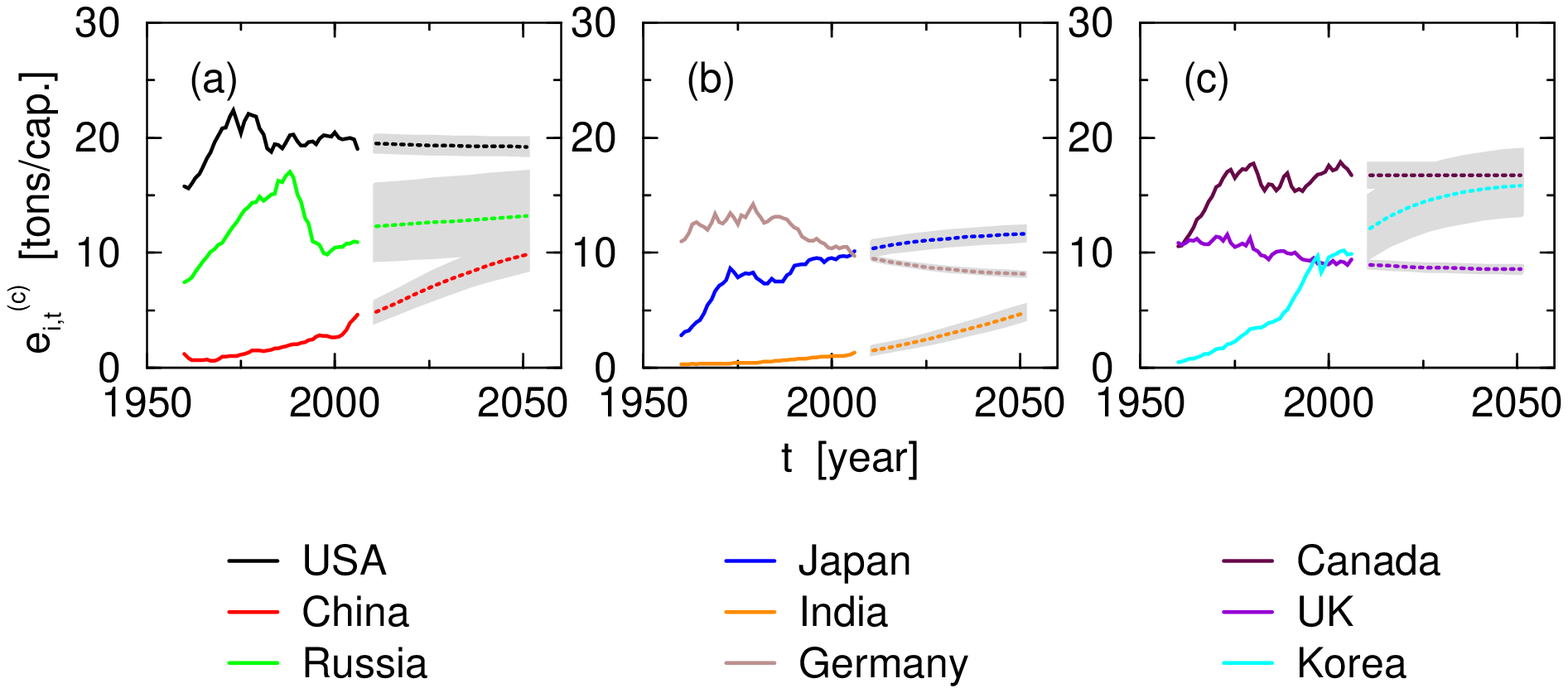}
\caption{
\label{fig:top9tot}
Examples of extrapolated CO$_2$ emissions per capita.
For the countries with top total emissions in $2000$,
we plot the measured values (solid lines) and
extrapolated values up to the middle of the 21st century (dotted lines).
The gray uncertainty range is obtained by including the statistical errors
of the regressions (one standard deviation each).
}
\end{figure*}

We take advantage of these correlations and assume that the system is ergodic,
i.e. that the process over time and over the statistical ensemble are the same.
In other words, we assume that these correlations
[main text Fig.~1, Eq.~(\ref{eq:co2hdiexp})] also hold for each
country individually, and apply the exponential regression:
\begin{equation}
\label{eq:co2hdiexpc}
\tilde{e}^{(\rm c)}_{i,t} = {\rm e}^{h_{i} d_{i,t}+g_{i}}
\enspace .
\end{equation}
Thus, for each country, we obtain the parameters~$h_i$ and~$g_i$,
characterizing how its emissions per capita are related to its development
(or vice versa).
Note that while in Eq.~(\ref{eq:co2hdiexp}) the year~$t$ is fixed,
leading to the time-dependent parameters~$h_t$ and~$g_t$,
in Eq.~(\ref{eq:co2hdiexpc}) the country~$i$ is fixed,
leading to the country-dependent parameters~$h_i$ and~$g_i$.
This regression, Eq.~(\ref{eq:co2hdiexpc}), is applied to $121$~countries for
which sufficient data is available,
i.e. at least $4$~pairs $e^{(\rm c)}_{i,t}$ and $d_{i,t}$.
Based on extrapolated HDI values we then calculate the corresponding future 
emissions per capita estimates.
Figure~\ref{fig:top9tot} shows for $9$~examples the past and
extrapolated values of emissions per capita.


\subsubsection{Correlations between CO$_2$~emissions per capita and 
HDI components}
\label{ssubsec:components}


%
\begin{figure*}
\includegraphics[width=0.75\textwidth]{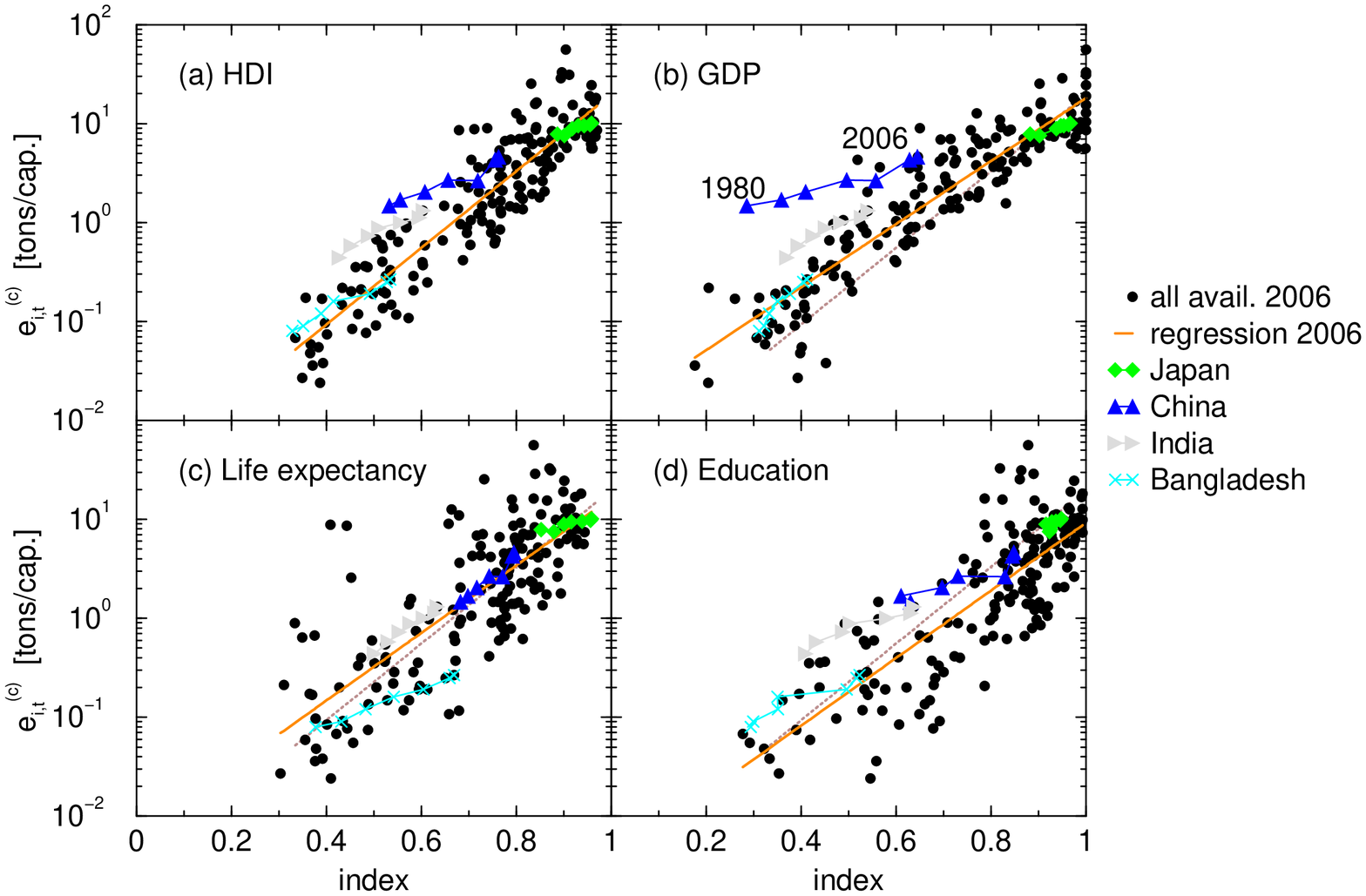}
\caption{
\label{fig:components}
Correlations between CO$_2$ emissions per capita and HDI 
as well as its components.
Panels~(a-d) are cross-plots in semi-logarithmic representation,
where each filled circle represents a country.
(a) depicts the CO$_2$ emissions per capita values versus the corresponding 
HDI values for the year 2006 ($172$ countries).
(b-d) depict the analogous for the HDI components, i.e. 
(b) GDP index, 
(c) life expectancy index, and 
(d) education index.
The slopes and correlation coefficients are listed in Tab.~\ref{tab:components}.
The Panels also include the trajectories (1980-2006) of 
Japan (green diamonds), 
China (blue triangle up), 
India (grey triangle right), and 
Bangladesh (cyan $\times$) 
evolving from the lower left to the upper right.
The solid straight lines are exponential fits, Eq.~(\ref{eq:co2hdiexp}), 
to the data and the dotted lines in (b-d) correspond to the fit from (a).
}
\end{figure*}
%


%
\begin{table}
\begin{tabular}{l|cc}
component & slope $h$ & corr. coeff. \\
\hline
HDI &	$8.93\pm 0.31$ &	$0.91$ \\
GDP &	$7.34\pm 0.24$ &	$0.92$ \\
life exp. &	$7.86\pm 0.48$ &	$0.78$ \\
education &	$7.87\pm 0.41$ &	$0.83$ \\
\end{tabular}

\caption{
\label{tab:components}
Slopes and correlation coefficients of the exponential fits, 
Eq.~(\ref{eq:co2hdiexp}), applied to the HDI and it's components.
}
\end{table}

In addition to the correlations between CO$_2$~emissions per capita and 
the HDI, we also calculated the correlations between CO$_2$ emissions per
capita and the three components of the HDI, 
i.e. a long and healthy life, knowledge, and a decent standard of living
(see Sec.~\ref{ssec:hdi}). 
As can be seen in Fig.~\ref{fig:components} for the year $2006$, 
in all cases we find clear correlations. 
In particular, we find that the slopes for the components are smaller 
than the one for HDI, see Tab.~\ref{tab:components}. 
This supports the usage of the HDI as summary measure.
However, the correlation coefficients of 
the life expectancy index vs. CO$_2$~emissions per capita and 
the education index vs. CO$_2$~emissions per capita 
are somewhat smaller ($0.78$ and $0.82$, respectively) than 
the one for the GDP index vs. CO$_2$~emissions per capita ($0.92$). 

By plotting the evolution of individual HDI components one can e.g. 
see that relative gains in education and life expectancy in Bangladesh 
supplant the gains in per capita GDP (Fig.~\ref{fig:components}). 
Obviously, the components them self are also correlated among each other 
(not shown).

\subsection{Estimating values for missing countries}
\label{subsec:extramis}

For $52$~countries out of $173$ the available data is not sufficient,
i.e. there are not enough values to perform the regressions
Eq.~(\ref{eq:hdilogreg}) or Eq.~(\ref{eq:co2hdiexpc}).
In order not to disregard these countries we take advantage of correlations,
i.e. countries with similar HDI have on average similar changes of
HDI as well as countries with similar emissions per capita have on
average similar changes of emissions.
In other words, in the $\ln e_{i,t}^{\rm (c)}$-$d_{i,t}$-plane,
the countries move similarly to their neighborhood.


Figure~3(c) and~(d) in the main text also shows how the regressions to the 
emissions per capita versus the HDI evolve.
The slope, $h_t$, becomes larger and the intercept, $g_t$, smaller.
In both cases the standard error remains approximately the same, showing
that the spreading of the cloud does not change.
In other words, if the countries would develop independently from each other,
then the error bars should increase with time.


%
\begin{figure*}
\includegraphics[width=0.7\textwidth]{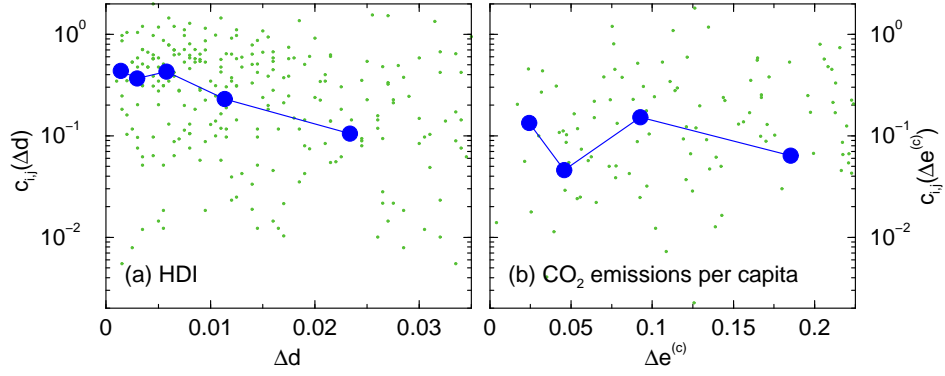}
\caption{
\label{fig:co2hdicorr}
Correlations of the changes in development and emissions per capita.
For observed data between the years~$2000$ and~$2005$,
we plot in (a) the correlation function, Eq.~(\ref{eq:hdicorr}),
of the temporal changes of the HDI as a function of the difference of the
countries in terms of HDI.
In (b) the analogue, namely the correlation function
of the temporal changes of the emissions per capita is plotted as a
function of the difference in terms of emissions per capita.
While the green dots represent the products of individual pairs,
the blue filled circles represent the averages in logarithmic bins.
}
\end{figure*}

In order to further support this feature,
in Fig.~\ref{fig:co2hdicorr} 
we show the correlations for both quantities.
Thus, for each pair of countries~$i$ and~$j$
(that are in the set with sufficient data),
we calculate
\begin{equation}
\label{eq:hdicorr}
c_{i,j}(\Delta d) =
\frac{(\delta d_i-\langle\delta d\rangle)(\delta d_j-\langle\delta d\rangle)}
{\sigma^2_{\delta d}}
\enspace ,
\end{equation}
where $\delta d_i=d_{i,2005}-d_{i,2000}$ is the difference in time,
$\langle\delta d\rangle$ is the average of $\delta d$ among all countries
providing enough data, and $\sigma^2_{\delta d}$ is the corresponding variance.
In Fig.~\ref{fig:co2hdicorr}(a) $c_{i,j}$ is plotted against
$\Delta d_{i,j}=|d_{j,2000}-d_{i,2000}|$, the difference in space of the
considered pair of countries.
One can see that the correlations decay exponentially following
$\tilde{c}_{i,j}(\Delta d)\approx {\rm e}^{-67.8 \Delta d -0.66}$.
This indicates that countries that have similar HDI also develop similarly.


For the emissions per capita we perform the analogous analysis,
replacing $\delta d_{i}$ by $\delta \ln e_{i}^{\rm (c)}$
in Eq.~(\ref{eq:hdicorr}) and consequently in the quantities
$\langle \ln \delta e^{\rm (c)} \rangle$,
$\sigma^2_{\delta \ln e^{\rm (c)}}$, and
$\Delta \ln e_{i,j}^{\rm (c)}$.
In Fig.~\ref{fig:co2hdicorr}(b) we obtain similar
results as for the HDI.
For the emissions, the correlations decay as
$\tilde{c}_{i,j}(\Delta \ln e^{\rm (c)})\approx {\rm e}^{-1.82 \Delta \ln e^{\rm (c)}-2.27}$.
For both, $\delta d_i$ and $\delta \ln e^{\rm (c)}$, the correlations were
calculated between the years~$2000$ and~$2005$.

We take advantage of these correlations and utilize 
them to extrapolate
$d_{i,t}$ and $e^{\rm (c)}_{i,t}$ by using the estimated correlation
functions as weights.
The change in development of a country~$k$, belonging to the
set of countries without sufficient data, we calculate with
\begin{equation}
\delta d_{k} =
\frac{\sum_j \left( \tilde{c}_{k,j}(\Delta d) \delta d_{j} \right)}
{\sum_j \tilde{c}_{k,j}(\Delta d)}
\enspace ,
\end{equation}
where the index~$j$ runs over the set of countries with sufficient data.
Then, the HDI in the following time step is
\begin{equation}
d_{k,t+1}=d_{k,t}+\delta d_{k,t}
\enspace .
\end{equation}
The analogous procedure is performed for the emissions per capita.

The results are shown in Fig.~\ref{fig:co2hdiexpmis}.
For comparison, the panels~(a) and~(b)
show the measured values for the years~$2000$ and~$2006$.
The panels~(c) to~(g) exhibit the extrapolated values, whereas the
black dots belong to the set of countries with sufficient data
(only HDI-extrapolation and HDI-CO$_2$-correlations)
and the brownish dots belong to the set of countries without sufficient
data.
In sum we can extrapolate the emissions for $172$ countries
(for one there is no $2006$ emissions value).
For most countries we obtain reasonable estimations
(see also Sec.~\ref{subsec:limitations}).
Panels~(h) and~(g) show the corresponding parameters~$h_t$ and~$g_t$
(slope and intercept).
The extrapolated values follow the tendency of the values for the past,
supporting the plausibility of this approach.
Nevertheless, the standard errors increase slightly in time, 
which indicates that the cloud of dots becomes slightly more disperse,
i.e. weaker ensemble correlations between~$e^{\rm (c)}_{i,t}$
and~$d_{i,t}$.

%
\begin{figure}
\includegraphics[width=0.80\columnwidth]{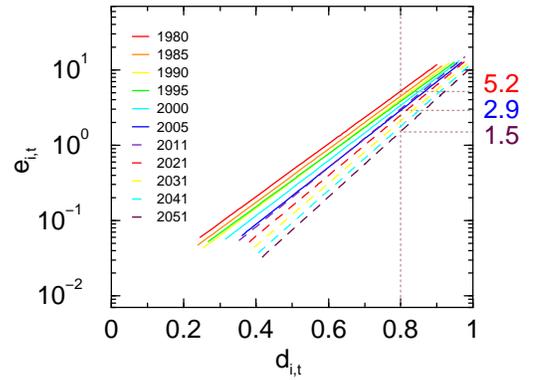}
\caption{
\label{fig:co2hdifits}
Evolving correlations between CO$_2$ emissions per capita and HDI.
The lines represent the linear regressions applied to the logarithm of
CO$_2$ emissions per capita versus HDI for the past (solid lines)
and our projections (dashed lines).
The numbers at the right edge correspond to the $e_{i,t}$ for which
the regressions cross $d_{i,t}=0.8$ in $1980$, $2005$ and projected for~$2051$.
}
\end{figure}
%

Figure~\ref{fig:co2hdifits} summarizes how the regressions
-- Eq.~(\ref{eq:co2hdiexp}) to the cloud of points representing
the countries -- evolve in the past and according to our projections.
Since the countries develop, the regression line moves towards larger values
of the HDI and at the same time its slope becomes steeper.
As a consequence, on average the per capita emissions of countries with
$d_{i,t}\simeq 0.8$ decrease with time
from approx.~$5.2$\,tons per year in $1980$
to approx.~$2.9$\,tons per year in $2005$ and
we expect it to reach approx.~$1.5$\,tons per year in $2051$.
This is in line with previous analysis \cite{SteinbergerR2010}.

\subsection{Uncertainty}
\label{subsec:uncertainty}

In order to obtain an uncertainty estimate of our projections,
we take into account the residuals of the regressions to the HDI versus time
and CO$_2$ versus HDI.
Thus, we calculate the root mean square deviations,
$\sigma^{(d)}_i$ and $\sigma^{(e)}_i$, respectively.
The upper and lower estimates of emissions per capita are then obtained
from
\begin{equation}
\tilde{d}_{i,t,(\pm)} = \frac{1}{1+{\rm e}^{-a_it+b_i \mp \sigma^{(d)}_i}}
\end{equation}
and
\begin{equation}
\tilde{e}^{(\rm c)}_{i,t,(\pm)} =
g_{i}{\rm e}^{h_{i} \tilde{d}_{i,t,(\pm)} \pm \sigma^{(e)}_i}
\enspace .
\end{equation}
In a rough approximation, assuming independence of the deviations,
the upper and lower bounds correspond to the range enclosing $90$\%.


%
\begin{figure*}
\includegraphics[width=0.7\textwidth]{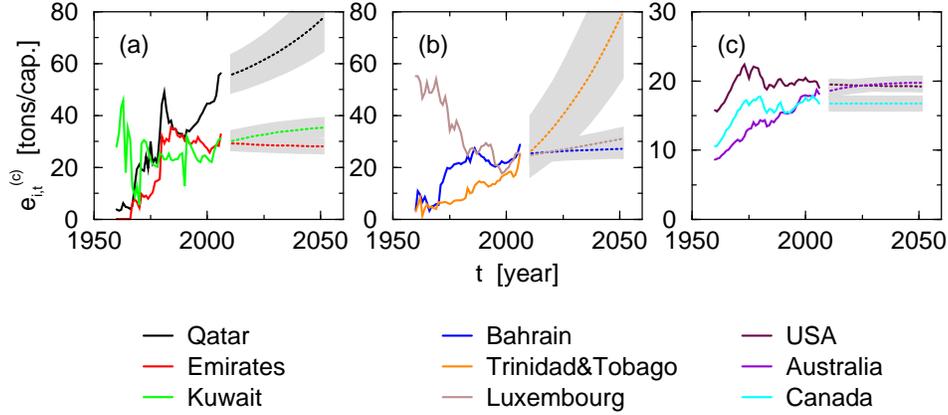}
\caption{
\label{fig:top9}
Examples of extrapolated CO$_2$ emissions per capita.
For the countries with top emissions per capita in $2006$,
we plot the measured values (solid lines) and extrapolated values up to
the middle of the 21st century (dotted lines).
Qatar and Luxembourg belong to those countries, which due to missing data
have been extrapolated utilizing correlations in the changes of 
$d_{i,t}$ as well as $e^{(c)}_{i,t}$ (Sec.~\ref{subsec:extramis}).
The gray uncertainty range is obtained by including the statistical errors
of the regressions (one standard deviation each).
Analogous to Fig.~\ref{fig:top9tot} but for different countries.
}
\end{figure*}

The obtained ranges can be seen as gray bands in 
Fig.~\ref{fig:top9tot} and~\ref{fig:top9}.
We find that the global cumulative CO$_2$ emissions 
between the years~$2000$ and~$2050$
discussed in the main text exhibit an uncertainty of
approx.~$12$\% compared to the typical value, 
which also includes uncertainty due to the population scenarios
(see Sec.~\ref{ssec:datapop} and Fig.~4 in the main text).

The global emissions we calculate for the years~$2000$ and~$2005$
(i.e. multiplying recorded CO$_2$ emissions per capita with
recorded population numbers, see Sec.~\ref{sec:cumemi})
are by less than $2$\% smaller than those provided by the WRI \cite{co2data}.
This difference, which can be understood as a systematic error,
can have two origins.
(i) Some countries are still missing. Either they are not included
in the data at all, or they cannot be considered, such as when we
multiply emissions per capita with the corresponding population and 
the two sets of countries do not match.
(ii) The population numbers we use might differ from the ones WRI uses.

\subsection{Limitations}
\label{subsec:limitations}


Since countries with already large HDI can only have small changes in
$d_{i,t}$, the emissions per capita also do not change much.
For example, for Australia, Canada, Japan, and the USA we obtain rather
stable extrapolations (Fig.~\ref{fig:top9tot} and~\ref{fig:top9}).
This could be explained by the large economies and the inertia they
comprise.
In contrast, for some countries with comparably small populations,
the extrapolated values of emissions per capita reach unreasonably high
values, such as for Qatar or Trinidad\&Tobago in Fig.~\ref{fig:top9}.


%
\begin{figure}
\includegraphics[width=0.9\columnwidth]{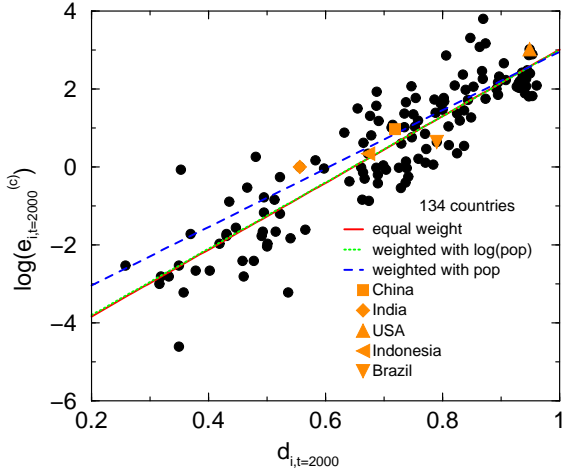}
\caption{
\label{fig:polyfitw}
Correlations between CO$_2$ emissions per capita and HDI.
For the year $2000$ three different ways of performing a regressions
are exemplified.
Solid line in the background: the regression when each country has the
same weight.
Dotted line: the countries have weights according to the logarithm of
their population.
Dashed line: the countries have weights according to their population.
For comparison the five most populous countries are highlighted.
}
\end{figure}

Since one may argue that countries with large populations 
should have more weight \cite{SteinbergerR2010} 
when fitting the per capita emissions 
versus the HDI, Eq.~(\ref{eq:co2hdiexp}),
in Fig.~\ref{fig:polyfitw},
for the year~$2000$, we employ three ways of
weighing.
While the solid line is the fit where all countries 
have the same weight, 
the dotted line is a regression where the points are weighted with
the logarithm of the country's population.
We found that it is almost identical to the unweighted one.
In contrast, the dashed line is a regression where the points are weighted
with the population of the countries
(not their logarithm as before).
The obtained regression differs from the other ones and as expected it is
dominated by the largest countries
(five of them are indicated in Fig.~\ref{fig:polyfitw}).
However, this difference does not influence our extrapolations
since we do not use the ensemble fit, Eq.~(\ref{eq:co2hdiexp}),
but instead regressions for individual countries, Eq.~(\ref{eq:co2hdiexpc}).

\subsection{Enhanced development approach}
\label{subsec:fastdevelop}

In addition to the DAU approach,
we also tested one of enhanced development where we
force the countries with $d<0.8$ to reach an HDI of $0.8$ by $2051$.
This can be done by parameterizing the HDI-regression through two points,
namely~$d_{i,2006}$ and~$d_{i,2050}=0.8$, instead of fitting
Eq.~(\ref{eq:hdilogreg}).
The corresponding emission values can then be estimated by
following the ensemble fit, Eq.~(\ref{eq:co2hdiexp}).
Nevertheless, since the relevant countries are rather small in
population and still remain with comparably small per capita emissions,
the difference in global emissions is minor,
namely at most an additional $3$\%
(cumulative emissions until $2050$, GO population scenario).
Thus, we do not further consider this enhanced development.

\section{Cumulative emissions}
\label{sec:cumemi}

To obtain the cumulative emission values,
shown in Fig.~2 of the main text,
we perform the following steps:
\begin{enumerate}
\item
Estimate the emissions per capita, $e^{\rm (c)}_{i,t}$,
according to the descriptions in Sec.~\ref{sec:co2project}.
\item\label{it:step2}
Multiply the per capita emissions with the population of the
corresponding countries, $e_{i,t}=e^{\rm (c)}_{i,t}\,p_{i,t}$,
resulting in the total annual emissions of each country.
\item
Calculate the cumulative emissions by
integrating the annual emission values,
$E_{i,t}=\sum_{\tau=t_0}^{t}e_{i,\tau}$, where we choose $t_0=2000$.
\end{enumerate}

The intersection of the set of countries with projected per capita emission
values with the set of countries with projected population values
consists of $165$~countries.

\section{Reduction scheme}
\label{sec:redsch}

In the main text we propose a CO$_2$-reduction scheme which is in line 
with our results. 
The reduction rate of the individual countries should depend on their 
individual HDI value. 
Thus, a country~$i$ reduces it's per capita emissions at year~$t$ 
according to 
\begin{equation}
\label{eq:co2red}
e^{\rm (c)}_{i,t-5\text{y}}\enspace\rightarrow\enspace (1-r_{i,t})\,e^{\rm (c)}_{i,t}
\end{equation} 
with the $5$-year reduction rate, $r_{i,t}$, which depends on the 
country's HDI following 
\begin{equation}
\label{eq:co2redux}
r_{i,t} = f\,(d_{i,t}-d^{*}) \quad \text{for} \quad d_{i,t}>d^{*}
\enspace ,
\end{equation}
involving two parameters, the development threshold, $0<d^{*}<1$, and 
the proportionality constant, $f>0$. 
The former determines at which HDI the countries start their reduction 
of per capita CO$_2$ emissions and the latter determines how strong the 
reduction rate increases with increasing HDI. 

Obviously, the larger $d^*$ is, the larger $f$ needs to be (and vice verse) 
so that global emissions can be limited.
Choosing the development threshold, $d^{*}=0.8$, we estimate 
that $f\simeq 3.3$ would lead to global cumulative emissions ranging 
between $850$ and $1100$\,Gt of CO$_2$ by $2050$ if reduction starts in $2015$ 
(assuming the same uncertainty as in DAU). 

Naturally, larger values of $f$ lead to smaller global emissions 
($f\simeq 3.3$ is a lower bound). 
However, the response is non-linear: 
$d^*=0.7$ requires $f\simeq 1.1$ and 
$d^*=0.6$ only $f\simeq 0.6$. 
For $d^*>0.8$ the emissions can practically not be restricted to the 
limit of $1000$\,Gt global emissions by $2050$ within the proposed 
reduction framework.

\section*{Acknowledgments}
The authors acknowledge the financial support from BaltCICA (Baltic Sea Region
Programme 2007-2013). They wish to thank the Federal Ministry for the
Environment, Nature Conservation, and Nuclear Safety of Germany who supports
this work within the framework of the International Climate Protection
Initiative.
We thank M. Boettle, S. Havlin, A. Holsten, D. Reusser, H.D. Rozenfeld, J.
Sehring, and J. Werg for discussions and comments. Furthermore, we thank A.
Schlums for help with the manuscript. The main author further thanks N.
Kozhevnikova for her refined sense of critique. All authors express their
gratitude for the Editor comments that largely benefited the current manuscript.




\end{document}